\documentclass[10pt,a4paper]{article}
\usepackage{amsmath}
\usepackage{amssymb}
\usepackage{amsfonts}
\usepackage{amstext}
\usepackage{amsbsy}
\usepackage[mathscr]{eucal}
\usepackage{graphicx}
\usepackage{color}
\usepackage[all]{xy}
\newcommand{\beq}{\begin{equation}}
\newcommand{\eeq}{\end{equation}}
\newcommand{\beqa}{\begin{eqnarray}}
\newcommand{\eeqa}{\end{eqnarray}}
\newcommand{\ket}[1]{| #1 \rangle}

\makeindex
\title{\Large\textbf{Noncommutative geometrical structures of entangled quantum states}}

\author{\textit{ Hoshang Heydari}\\
        \small\textit{Physics Department, Stockholm university 10691 Stockholm Sweden}\\
\\\small\textit{Email: hoshang@fysik.su.se}}

\date{}
\begin{document}

\maketitle \thispagestyle{empty}

\maketitle
\begin{abstract}
We study the noncommutative geometrical structures of quantum entangled states. We show that the space of a pure entangled state is a  noncommutative space. In particular we show that by rewritten the conifold or the Segre variety we can get a $q$-deformed relation in noncommutative geometry. We generalized our construction into a multi-qubit state. We also in detail discuss the noncommutative geometrical structure of a three-qubit state.
\end{abstract}

\section{Introduction}
Quantum entangled states are the main resources in the field of quantum information science.
These states also have very rich geometrical and topological structures. Geometrically the space of a pure quantum state  is a complex projective space, that is $\mathcal{PH}=\mathcal{H}/\sim$, where $\mathcal{H}$ is the Hilbert space and $\sim$ is a equivalence relation. For example, if we let $\mathcal{H}=\mathbb{C}^{n+1}$, then $\mathcal{PH}=\mathbb{CP}^{n}$. Recently, we also have established relation between multi-projective variety (space) and pure quantum multipartite state. We have shown that the multi-projective Segre variety is the space of separable quantum composite systems and so it can distinguish between separable and entangled multipartite quantum systems \cite{Hosh5}. Topologically, the space of two level state or a qubit can be described by Block sphere $S^{2}\cong\mathbb{CP}^{1}$ which is obtained by Hopf fibration $\mathbb{S}^{1}\hookrightarrow \mathbb{S}^{3}\longrightarrow \mathbb{S}^{2}$. Generally, we have the following Hopf fibration $S^{2n+1}\longrightarrow \mathbb{CP}^{n}$. We also shown that there is relation between Hopf fibration and multi-qubit states \cite{Hosh6}.
For a pure multi-qubit state
\begin{eqnarray}\ket{\Psi}&=&\sum^{1}_{x_{m}=0}\sum^{1}_{x_{m-1}=0}\cdots
\sum^{1}_{
x_{1}=0}\alpha_{x_{m}x_{m-1}\cdots x_{1}}\ket{x_{m}x_{m-1}\cdots
x_{1}},\\\nonumber&=&
\sum^{1}_{x_{m}=0}\sum^{1}_{x_{m-1}=0}\cdots
\sum^{1}_{
x_{1}=0}\alpha_{x}\ket{x},\
\end{eqnarray}
where  $\ket{x_{m}x_{m-1}\cdots
x_{1}}=\ket{x_{m}}\otimes\ket{x_{m-1}}\otimes\cdots\otimes\ket{x_{1}}$ are orthonormal basis in $
\mathcal{H}_{\mathcal{Q}}=\mathcal{H}_{\mathcal{Q}_{1}}\otimes
\mathcal{H}_{\mathcal{Q}_{2}}\otimes\cdots\otimes\mathcal{H}_{\mathcal{Q}_{m}}
$ and $x=x_{m-1}2^{m-1}+x_{m-2}2^{m-2}+\cdots+x_{0}2^{0}$, the set of state is defined by
$\mathcal{SH}_{\mathcal{Q}}=\{\ket{\Psi}\in\mathcal{H}_{\mathcal{Q}}:\langle\Psi\ket{\Psi}=1\}$.
In this paper, we establish a relation between noncommutative geometry and quantum entangled state. In particular, we show that by resolving the singularity of conifold we get a space which can be written  in such form that  is a $q$-deformed relation in noncommutative geometry. In section \ref{sec2} we give a short introduction to multi-projective variety and in section \ref{sec3} we review the construction of conifold and the quantum plane. In section \label{sec5} we establish our first result, namely, the noncommutative structure two-qubit state. Finally, in section \label{sec6} we generalize our result into multi-qubit state and we also discuss the three-qubit state as an illustrative example.

\section{Multi-projective  variety} \label{sec2}
In this section, we will review the construction of projective variety and in particular the multi-projective
Segre variety.
Here are some prerequisites on projective algebraic geometry
\cite{Griff78,Mum76}.
Let $\mathbb{C}$ be a complex algebraic field. Then, an affine
$n$-space over $\mathbb{C}$ denoted $\mathbb{C}^{n}$ is the set of
all $n$-tuples of elements of $\mathbb{C}$. An element
$P\in\mathbb{C}^{n}$ is called a point of $\mathbb{C}^{n}$ and if
$P=(a_{1},a_{2},\ldots,a_{n})$ with $a_{j}\in\mathbb{C}$, then
$a_{j}$ is called the coordinates of $P$.

Let $\mathbb{C}[z]=\mathbb{C}[z_{1},z_{2}, \ldots,z_{n}]$ denotes the polynomial
algebra in $n$  variables with complex coefficients. Then, given a
set of $q$ polynomials $\{g_{1},g_{2},\ldots,g_{q}\}$ with $g_{i}\in
\mathbb{C}[z]$, we define a complex affine variety as
\begin{eqnarray}
&&\mathcal{V}_{\mathbb{C}}(g_{1},g_{2},\ldots,g_{q})=\{P\in\mathbb{C}^{n}:
g_{i}(P)=0~\forall~1\leq i\leq q\},
\end{eqnarray}
A complex projective space $\mathbb{CP}^{n}$ is defined to be the
set of lines through the origin in $\mathbb{C}^{n+1}$, that is,
\begin{equation}
\mathbb{CP}^{n}=\frac{\mathbb{C}^{n+1}-{0}}{
(u_{1},\ldots,u_{n+1})\sim(v_{1},\ldots,v_{n+1})},~\lambda\in
\mathbb{C}-0,~v_{i}=\lambda u_{i} ~\forall ~0\leq i\leq n+1.
\end{equation}
Given a set of homogeneous polynomials
$\{g_{1},g_{2},\ldots,g_{q}\}$  with $g_{i}\in \mathbb{C}[z]$, we define a
complex projective variety as
\begin{eqnarray}
&&\mathcal{V}(g_{1},\ldots,g_{q})=\{O\in\mathbb{CP}^{n}:
g_{i}(O)=0~\forall~1\leq i\leq q\},
\end{eqnarray}
where $O=[a_{1},a_{2},\ldots,a_{n+1}]$ denotes the equivalent class
of point $\{\alpha_{1},\alpha_{2},\ldots,$
$\alpha_{n+1}\}\in\mathbb{C}^{n+1}$. We can view the affine complex
variety
$\mathcal{V}_{\mathbb{C}}(g_{1},g_{2},\ldots,g_{q})\subset\mathbb{C}^{n+1}$
as a complex cone over the complex projective variety
$\mathcal{V}(g_{1},g_{2},\ldots,g_{q})$.

  We can map the
product of  spaces $\mathbb{CP}^{1}\times\mathbb{CP}^{1}
\times\cdots\times\mathbb{CP}^{1}$ into a projective space by
its Segre embedding as follows. Let
$(\alpha^{i}_{0},\alpha^{i}_{1})$  be
points defined on the $i$th complex projective space
$\mathbb{CP}^{1}$. Then the Segre map is given by
\begin{equation}
\begin{array}{ccc}
  \mathcal{S}_{2,\ldots,2}:\mathbb{CP}^{1}\times\mathbb{CP}^{1}
\times\cdots\times\mathbb{CP}^{1}&\longrightarrow&
\mathbb{CP}^{2^{m}-1}\\
 ((\alpha^{1}_{0},\alpha^{1}_{1}),\ldots,
 (\alpha^{m}_{0},\alpha^{m}_{1})) & \longmapsto&
 (\alpha^{m}_{i_{m}}\alpha^{m-1}_{i_{m-1}}\cdots\alpha^{1}_{i_{1}}). \\
\end{array}
\end{equation}
Now, let $\alpha_{i_{m}i_{m-1}\cdots i_{1}}$,$0\leq i_{s}\leq 1$
be a homogeneous coordinate-function on
$\mathbb{CP}^{2^{m}-1}$. Moreover, let us consider
a multi-qubit quantum system
 and let
$
\mathcal{A}=\left(\alpha_{i_{m}i_{m-1}\ldots i_{1}}\right)_{0\leq
i_{s}\leq 1},
$
for all $j=1,2,\ldots,m$. $\mathcal{A}$ can be realized as the
following set $\{(i_{1},i_{2},\ldots,i_{m}):1\leq i_{s}\leq
2,\forall~s\}$, in which each point $(i_{m},i_{m-1},\ldots,i_{1})$
is assigned the value $\alpha_{i_{m}i_{m-1}\ldots i_{1}}$. This
realization of $\mathcal{A}$ is called an $m$-dimensional box-shape
matrix of size $2\times2\times\cdots\times 2$, where we
associate to each such matrix a sub-ring
$\mathrm{S}_{\mathcal{A}}=\mathbb{C}[\mathcal{A}]\subset\mathrm{S}$,
where $\mathrm{S}$ is a commutative ring over the complex number
field. For each $s=1,2,\ldots,m$, a two-by-two minor about the
$j$-th coordinate of $\mathcal{A}$ is given by
\begin{eqnarray}\label{segreply1}
&&\mathcal{P}^{s}_{x_{m}y_{m};x_{m-1}y_{m-1};\ldots;x_{1}y_{1}}=
\alpha_{x_{m}x_{m-1}\ldots x_{1}}\alpha_{y_{m}y_{m-1}\ldots y_{1}}
\\\nonumber&&-
\alpha_{x_{m}x_{m-1}\ldots x_{s+1}y_{s}x_{s-1}\ldots
x_{1}}\alpha_{y_{m}y_{m-1} \ldots y_{s+1} x_{s}y_{s-1}\ldots y_{m}}\in
\mathrm{S}_{\mathcal{A}}.
\end{eqnarray}
Then the ideal $\mathcal{I}^{m}_{\mathcal{A}}$ of
$\mathrm{S}_{\mathcal{A}}$ is generated by
$\mathcal{P}^{s}_{x_{m}y_{m};x_{m-1}y_{m-1};\ldots;x_{1}y_{1}}$  and
describes the separable states in $\mathbb{CP}^{2^{m}-1}$. The image of the Segre embedding
$\mathrm{Im}(\mathcal{S}_{2,2,\ldots,2})$, which again
is an intersection of families of quadric hypersurfaces in
$\mathbb{CP}^{2^{m}-1}$, is called Segre variety
and it is given by
\begin{eqnarray}\label{eq: submeasure}
\mathrm{Im}(\mathcal{S}_{2,2,\ldots,2})&=&\bigcap_{\forall
s}\mathcal{V}\left(\mathcal{P}^{s}_{x_{m}y_{m};x_{m-1}y_{m-1};\ldots;x_{1}y_{1}}\right).
\end{eqnarray}
In the following section we establish relations between deformed Segre variety and $q$-deformed
noncommutative geometry.

\section{Conifold and quantum plane}\label{sec3}
In this section we will give a short review of conifold and quantum plane.
An
example of real (complex) affine variety is conifold which is
defined by
\begin{equation}
\mathcal{V}_{\mathbb{C}}(z)=\{(z_{1},z_{2},z_{3},z_{4})
\in\mathbb{C}^{4}: \sum^{4}_{i=1}z^{2}_{i}=0\}.
\end{equation}
and conifold as a real affine variety is define by
\begin{equation}
\mathcal{V}_{\mathbb{R}}(f_{1},f_{2})=\{(u_{1},\ldots,u_{4},v_{1},\ldots,v_{4})\in\mathbb{R}^{8}:
\sum^{4}_{i=1}u^{2}_{i}=\sum^{4}_{j=1}v^{2}_{j},\sum^{4}_{i=1}u_{i}v_{i}=0
\}.
\end{equation}
where $f_{1}=\sum^{4}_{i=1}(u^{2}_{i}-v^{2}_{i})$ and
$f_{2}=\sum^{4}_{i=1}u_{i}v_{i}$. This can be seen by defining
$z=u+iv$ and identifying imaginary and real part of equation
$\sum^{4}_{i=1}z^{2}_{i}=0$.  As a real space, the conifold
is cone in $\mathbb{R}^{8}$ with top the origin and base space the
compact manifold $\mathbb{S}^{2}\times\mathbb{S}^{3}$.
  One can reformulate this relation in
term of a theorem. The conifold $
\mathcal{V}_{\mathbb{C}}(\sum^{4}_{i=1}z^{2}_{i}) $ is the complex
cone over the Segre variety $\mathbb{CP}^{1}
  \times\mathbb{CP}^{1}\longrightarrow\mathbb{CP}^{3}$. To see this let us make a complex linear
  change of coordinate
 \begin{equation}
 \left(
   \begin{array}{cc}
    \alpha^{'}_{00} & \alpha^{'}_{01} \\
     \alpha^{'}_{10} & \alpha^{'}_{11}\\
   \end{array}
 \right)\longrightarrow \left(
   \begin{array}{cc}
    z_{1}+iz_{2} & -z_{4}+iz_{3} \\
    z_{4}+iz_{3} & z_{1}-iz_{2}\\
   \end{array}
 \right).
 \end{equation}
    Thus after this linear
  coordinate transformation we have
  \begin{equation}\label{Conifold}
    \mathcal{V}_{\mathbb{C}}(\alpha^{'}_{00}\alpha^{'}_{11}-\alpha^{'}_{01}\alpha^{'}_{10})
    =\mathcal{V}_{\mathbb{C}}(\sum^{4}_{i=1}z^{2}_{i})\subset\mathbb{C}^{4}.
\end{equation}
Thus we can think of conifold as a complex cone over $\mathbb{CP}^{1}
  \times\mathbb{CP}^{1}$. Moreover, we can remove the singularity of  complex conifold $T^{*}\mathbb{S}^{3}$ with a global complex deformation parameter $\Omega$. In this case we have  a hypersurface $H=H(\alpha_{0},\alpha_{1},\alpha_{2},\alpha_{3})$ which is embedded in $\mathbb{C}^{4}$ by \begin{equation}
 \alpha_{00}\alpha_{11}-\alpha_{01}\alpha_{10}= \alpha_{0}\alpha_{3}-\alpha_{1}\alpha_{2}=\Omega.
\end{equation}
We will return to this equation in the following sections when we discuss noncommutative geometrical structures of two-qubits.

Next we will give a short introduction to quantum plane \cite{kassel}. Let $\mathbb{C}$ be a complex number field and $q$ be a invertible element of $\mathbb{C}$. Moreover, let $I_{q}$ be the two side ideal of the free algebra $\mathbb{C}\{u,v\}$ which is generated by $vu-quv$. Then
 the quantum plane is defined to be the quotient-algebra
  \begin{equation}
\mathbb{C}_{q}[u,v]=\mathbb{C}\{u,v\}/I_{q}.
\end{equation}
The quantum plane is non-commutative if $q\neq1$. The ideal $I_{q}$ is generated by homogeneous degree two element. For any pair $(i,j)$, we have $v^{j}u^{i}-qu^{i}v^{j}=0$ and Given  any $\mathbb{C}$-algebra $R$, there is a bijection
\begin{equation}
\mathrm{Hom}(\mathbb{C}_{q}[u,v],R)\cong\{(U,V)\in R\times R:VU-qUV=0\}.
\end{equation}
The pair $(U,V)$ satisfying above relation are called a $R$-point of the quantum plane.  There is direct relation between our construction in the following section and quantum plane.
Our short review of quantum plane could be  important in future investigation of quantum geometry and quantum entangled states.

\section{Noncommutative geometrical structure of  two-qubits}\label{sec5}
In this section we investigate a pure two-qubit state based on noncommutative geometry.
A pure two-qubit state is given by
 \begin{equation}
 \ket{\Psi}=\alpha_{00}\ket{00}+\alpha_{01}\ket{01}
+\alpha_{10}\ket{10}+\alpha_{11}\ket{11}.
\end{equation}
Now, based on the Segre variety construction, the separable state of such two-qubit state is given by
\begin{equation}
\alpha_{00}\alpha_{11}-\alpha_{01}\alpha_{10}=0
\end{equation}
Thus, for entangled state we have $\alpha_{00}\alpha_{11}-\alpha_{01}\alpha_{10}\neq0$. As we have discussed
 this condition is also related to deformation of conifold. Now, let
 \begin{equation}
 \left(
   \begin{array}{cc}
     \alpha_{00} & \alpha_{01} \\
     \alpha_{10} & \alpha_{11} \\
   \end{array}
 \right)=\left(
   \begin{array}{cc}
     \alpha_{0} & \alpha_{1} \\
     \alpha_{2} & \alpha_{3} \\
   \end{array}
 \right)=\left(
           \begin{array}{cc}
             u_{1} & u_{2} \\
              v_{1} & v_{2} \\
           \end{array}
         \right).
\end{equation}
Then, for this deformed variety we have
 \begin{equation}
\alpha_{00}\alpha_{11}-\alpha_{01}\alpha_{10}=u_{1}v_{2}-u_{2}v_{1}=\Omega
\end{equation}
which in this form  is a $q$-deformed relation in noncommutative geometry \cite{Closset}. Now, let $\mu_{i}=(u_{1},u_{2})$ and $\nu_{i}=(v_{1},v_{2})$. We can also write this equations as
$\varepsilon^{ij}\mu_{i}\nu_{j}=\Omega$, $\varepsilon^{ij}\mu_{i}\mu_{j}=0$, and
$\varepsilon^{ij}\nu_{i}\nu_{j}=0$, where $\varepsilon^{ij}$ is an antisymmetric tensor. Note that  the  relation $\varepsilon^{ij}\mu_{i}\nu_{j}=\Omega$ express $SL(2,\mathbb{C})$ invariance of conifold hypersurface $H$ in complex space $\mathbb{C}^{4}$.
 We also can write these equations as
 \begin{equation}
\mu_{[i}\nu_{j]}=\Phi_{ij},~~\mu_{[i}\nu_{j]}=0~,~~\text{and}~~
 \nu_{[i}\nu_{j]}=0,
 \end{equation}
  where $\Phi_{ij}=\varepsilon_{ij}\Omega/2$.
 Next, we set $\mu_{i}=\Lambda_{1i}$ and $\nu_{i}=\Lambda_{2i}$, then we get
  \begin{equation}
\Lambda_{ki}\Lambda_{lj}-\Lambda_{kj}\Lambda_{li}=\varepsilon_{kl}\Phi_{ij}
=\Lambda_{ki}\Lambda_{lj}-\mathfrak{R}^{mn}_{kl}\Lambda_{mj}\Lambda_{ni},
\end{equation}
where $\mathfrak{R}^{mn}_{kl}=\varepsilon^{m}_{k}\varepsilon^{n}_{l}$. This is a noncommutative space of a pure two-qubit entangled state. Moreover, this rewriting of deformed
variety for two-qubit could alow us to borrow techniques and tools from
 theory of $q$-deformed noncommutative geometry to investigate the structures of entangled states.
\section{Multi-qubit states}\label{sec6}
One would now ask if it possible to establish relation between multipartite quantum states and noncommutative geometry. The answer seems to be positive, since based on the multi-projective Segre variety, the complectly separable set of pure state is give by quadratic polynomial defined by equation (\ref{segreply1}).
But, when discussing a multi-qubit state it is better to consider the Segre ideals,
 \begin{equation}
\begin{array}{ccc}
   \left(
   \begin{array}{cccc}
     \alpha_{00\cdots00} & \alpha_{00\cdots01}&\cdots & \alpha_{01\cdots11} \\
     \alpha_{10\cdots00} & \alpha_{10\cdots01} &\cdots & \alpha_{11\cdots11}\\
   \end{array}
 \right)&=&\left(
           \begin{array}{cccc}
             u^{p_{1}}_{1} & u^{p_{1}}_{2} & \cdots& u^{p_{1}}_{2^{m-1}}\\
              v^{p_{1}}_{1} & v^{p_{1}}_{2}& \cdots & v^{p_{1}}_{2^{m-1}} \\
           \end{array}
         \right), \\
  \left(
   \begin{array}{cccc}
     \alpha_{00\cdots00} & \alpha_{00\cdots01}&\cdots & \alpha_{101\cdots1} \\
     \alpha_{010\cdots0} & \alpha_{01\cdots01} &\cdots & \alpha_{11\cdots11}\\
   \end{array}
 \right)&=&\left(
           \begin{array}{cccc}
             u^{p_{2}}_{1} & u^{p_{2}}_{2} & u^{p_{2}}_{3}& u^{p_{2}}_{2^{m-1}}\\
              v^{p_{2}}_{1} & v^{p_{2}}_{2}& v^{p_{2}}_{3} & v^{p_{2}}_{2^{m-1}} \\
           \end{array}
         \right),\\
  &\vdots& \\
  \left(
   \begin{array}{cccc}
     \alpha_{00\cdots00} & \alpha_{0\cdots010}&\cdots & \alpha_{11\cdots10} \\
     \alpha_{00\cdots01} & \alpha_{0\cdots011} &\cdots & \alpha_{11\cdots11}\\
   \end{array}
 \right)&=&\left(
           \begin{array}{cccc}
             u^{p_{m}}_{1} & u^{p_{m}}_{2} & \cdots& u^{p_{m}}_{2^{m-1}}\\
              v^{p_{m}}_{1} & v^{p_{m}}_{2}& \cdots & v^{p_{m}}_{2^{m-1}} \\
           \end{array}
         \right).
\end{array}
\end{equation}
Now we define
\begin{equation}\label{minors}
\mathrm{Minors}^{p_{s}}_{2\times2}\left(
           \begin{array}{cccc}
             u^{p_{s}}_{1} & u^{p_{s}}_{2} & \cdots& u^{p_{s}}_{2^{m-1}}\\
              v^{p_{s}}_{1} & v^{p_{s}}_{2}& \cdots & v^{p_{s}}_{2^{m-1}} \\
           \end{array}
         \right)=\Omega^{p_{s}},  ~~ s=1,2,\ldots,m
\end{equation}
where $p_{s}=\frac{2^{m-1}(2^{m-1}-1)}{2}$, is the number of quadratic polynomial defining the Segre variety of the multi-qubits and $\mathrm{Minors}^{p_{s}}_{2\times2}$ is
the $2\times2$ minors of the above $2\times2^{m-1}$ matrices.
Then,  for example a multi-qubit deformed variety is given by
 \begin{equation}
\alpha_{00\cdots0}\alpha_{10\cdots01}-\alpha_{0\cdots01}\alpha_{10\cdots00}=
u^{p_{1}}_{1}v^{p_{1}}_{2}-u^{p_{1}}_{2}v^{p_{1}}_{1}=\Omega^{p_{1}},~~\text{for}~~ p_{1}=1
\end{equation}
which in this form  is a $q$-deformed relation in noncommutative geometry, where $p_{s}$, for $s=1,2,\ldots,m$ is the number of quadratic polynomial defining the Segre variety of the multi-qubits. Now, let $\mu^{p_{s}}_{i_{s}}=(u^{p_{s}}_{1},u^{p_{s}}_{2},\ldots,u^{p_{s}}_{2^{m-1}})$ and $\nu^{p_{s}}_{i_{s}}=(v^{p_{s}}_{1},v^{p_{s}}_{2},\ldots,v^{p_{s}}_{2^{m-1}})$. Then, we can also write this equations as
$\varepsilon^{i_{s}j_{s}}\mu^{p_{s}}_{i_{s}}\nu^{p_{s}}_{j_{s}}=\Omega^{p_{s}}$, $\varepsilon^{i_{s}j_{s}}\mu^{p_{s}}_{i_{s}}\mu^{p_{s}}_{j_{s}}=0$, and
$\varepsilon^{i_{s}j_{s}}\nu^{p_{s}}_{i_{s}}\nu^{p_{s}}_{j_{s}}=0$
or as
 \begin{equation}
\mu^{p_{s}}_{[i_{s}}\nu^{p_{s}}_{j_{s}]}=\Phi^{p_{s}}_{i_{s}j_{s}},
~~\mu^{p_{s}}_{[i_{s}}\nu^{p_{s}}_{j_{s}]}=0~,~~\text{and}~~
 \nu^{p_{s}}_{[i_{s}}\nu^{p_{s}}_{j_{s}]}=0,
 \end{equation}
  where $\Phi^{p_{s}}_{i_{s}j_{s}}=\varepsilon_{i_{s}j_{s}}\Omega^{p_{s}}/2$.
 Next, following the same procedure, we let $\mu^{p_{s}}_{i_{s}}=\Lambda^{p_{s}}_{1i_{s}}$ and $\nu^{p_{s}}_{i_{s}}=\Lambda^{p_{s}}_{2i_{s}}$. Then we get
  \begin{equation}
\Lambda^{p_{s}}_{k_{s}i_{s}}\Lambda^{p_{s}}_{l_{s}j_{s}}-
\Lambda^{p_{s}}_{k_{s}j_{s}}\Lambda^{p_{s}}_{l_{s}i_{s}}=
\varepsilon^{p_{s}}_{k_{s}l_{s}}\Phi^{p_{s}}_{i_{s}j_{s}}
=\Lambda^{p_{s}}_{k_{s}i_{s}}\Lambda^{p_{s}}_{l_{s}j_{s}}-\mathfrak{R}^{m_{s}n_{s}}_{k_{s}l_{s}},
\Lambda^{p_{s}}_{m_{s}j_{s}}\Lambda^{p_{s}}_{n_{s}i_{s}},
\end{equation}
where $\mathfrak{R}^{m_{s}n_{s}}_{k_{s}l_{s}}=\varepsilon^{m_{s}}_{k_{s}}\varepsilon^{n_{s}}_{l_{s}}$.
To illustrate our construction we in detail discuss a three-qubit state  which  is given by $\ket{\Psi}=\sum^{1}_{x_{3},x_{2},x_{1}=0}
\alpha_{x_{3}x_{2}x_{1}} \ket{x_{3}x_{2}x_{1}}$.
Now, based on the Segre variety construction, the separable state of such three-qubit state is given by
\begin{eqnarray}\label{eq: submeasure}
\mathrm{Im}(\mathcal{S}_{2,2,2})&=&\bigcap_{\forall
s}\mathcal{V}\left(\mathcal{P}^{s}_{x_{3}y_{3};x_{2}y_{2};x_{1}y_{1}}\right).
\end{eqnarray}
Thus, for entangled state we have the following condition
\begin{equation}
\alpha_{x_{3}x_{2}x_{1}}\alpha_{y_{3}y_{2}y_{1}}-
\alpha_{x_{3}y_{s}
x_{1}}\alpha_{y_{3} x_{s} y_{3}}\neq0.
\end{equation}
 As we have discussed
 this condition is also related to deformation of conifold. Since we can apply the same procedure to establish relation between quantum entangled states and noncommutative geometry as  we have done for two-qubits. In this case we need to consider  all quadratic polynomials $\alpha_{x_{3}x_{2}x_{1}}\alpha_{y_{3}y_{2}y_{1}}-
\alpha_{x_{3}y_{s}
x_{1}}\alpha_{y_{3} x_{s} y_{3}}$. We can also consider the Segre ideals for three-qubits,
 \begin{equation}
 \left(
   \begin{array}{cccc}
     \alpha_{000} & \alpha_{001}&\alpha_{010} & \alpha_{011} \\
     \alpha_{100} & \alpha_{101} &\alpha_{110} & \alpha_{111}\\
   \end{array}
 \right)=\left(
           \begin{array}{cccc}
             u^{p_{1}}_{1} & u^{p_{1}}_{2} & u^{p_{1}}_{3}& u^{p_{1}}_{4}\\
              v^{p_{1}}_{1} & v^{p_{1}}_{2}& v^{p_{1}}_{3} & v^{p_{1}}_{4} \\
           \end{array}
         \right),
\end{equation}
 \begin{equation}
 \left(
   \begin{array}{cccc}
     \alpha_{000} & \alpha_{001}&\alpha_{100} & \alpha_{101} \\
     \alpha_{010} & \alpha_{011} &\alpha_{110} & \alpha_{111}\\
   \end{array}
 \right)=\left(
           \begin{array}{cccc}
             u^{p_{2}}_{1} & u^{p_{2}}_{2} & u^{p_{2}}_{3}& u^{p_{2}}_{4}\\
              v^{p_{2}}_{1} & v^{p_{2}}_{2}& v^{p_{2}}_{3} & v^{p_{2}}_{4} \\
           \end{array}
         \right),
\end{equation}
 \begin{equation}
 \left(
   \begin{array}{cccc}
     \alpha_{000} & \alpha_{100}&\alpha_{010} & \alpha_{110} \\
     \alpha_{001} & \alpha_{101} &\alpha_{011} & \alpha_{111}\\
   \end{array}
 \right)=\left(
           \begin{array}{cccc}
             u^{p_{3}}_{1} & u^{p_{3}}_{2} & u^{p_{3}}_{3}& u^{p_{3}}_{4}\\
              v^{p_{3}}_{1} & v^{p_{3}}_{2}& v^{p_{3}}_{3} & v^{p_{3}}_{4} \\
           \end{array}
         \right).
\end{equation}
Now the equation (\ref{minors}) for a three-qubit system takes the following form
\begin{equation}
\mathrm{Minors}^{p_{s}}_{2\times2}\left(
           \begin{array}{cccc}
             u^{p_{s}}_{1} & u^{p_{s}}_{2} & u^{p_{s}}_{3}& u^{p_{s}}_{4}\\
              v^{p_{s}}_{1} & v^{p_{s}}_{2}& v^{p_{s}}_{3} & v^{p_{s}}_{4} \\
           \end{array}
         \right)=\Omega^{p_{s}},  ~~ s=1,2,3,
\end{equation}
where $p_{s}=\frac{2^{3-1}(2^{3-1}-1)}{2}=6$, is the number of quadratic polynomial defining the Segre variety of the three-qubits and $\mathrm{Minors}^{p_{s}}_{2\times2}$ is the $2\times 2$ minors of the above $2\times2^{3-1}$ matrices.  In this form  we have again a $q$-deformed relation in noncommutative geometry.
For instance a deformed variety is given by
 \begin{equation}
\alpha_{000}\alpha_{101}-\alpha_{001}\alpha_{100}=u^{p_{1}}_{1}v^{p_{1}}_{2}-u^{p_{1}}_{2}v^{p_{1}}_{1}
=\Omega^{p_{1}},~~\text{for}~~ p_{1}=1.
\end{equation}
Now, let $\mu^{p_{s}}_{i_{s}}=(u^{p_{s}}_{1},u^{p_{s}}_{2},u^{p_{s}}_{3},u^{p_{s}}_{4})$ and $\nu^{p_{s}}_{i_{s}}=(v^{p_{s}}_{1},v^{p_{s}}_{2},v^{p_{s}}_{3},v^{p_{s}}_{4})$.  Then we have
$\varepsilon^{ij}\mu^{p_{s}}_{i_{s}}\nu^{p_{s}}_{j_{s}}=\Omega^{p_{s}}$, $\varepsilon^{ij}\mu^{p_{s}}_{i_{s}}\mu^{p_{s}}_{j_{s}}=0$, and
$\varepsilon^{ij}\nu^{p_{s}}_{i_{s}}\nu^{p_{s}}_{j_{s}}=0$.
 We can also  write these equations as
 \begin{equation}
\mu^{p_{s}}_{[i_{s}}\nu^{p_{s}}_{j]}=\Phi^{p_{s}}_{i_{s}j_{s}}=
\varepsilon_{i_{s}j_{s}}\Omega^{p_{s}}/2,~~\mu^{p_{s}}_{[i_{s}}\nu^{p_{s}}_{j_{s}]}=0~,~~\text{and}~~
 \nu^{p_{s}}_{[i_{s}}\nu^{p_{s}}_{j]}=0,
 \end{equation}
 If, we set $\mu^{p_{s}}_{i_{s}}=\Lambda^{p_{s}}_{1i_{s}}$ and $\nu^{p_{s}}_{i_{s}}=\Lambda^{p_{s}}_{2i_{s}}$, then we get the following set of $q$-deformed relations
\begin{equation}
\begin{array}{c}
  \Lambda^{p_{1}}_{k_{1}i_{1}}\Lambda^{p_{1}}_{l_{1}j_{1}}-
  \Lambda^{p_{1}}_{k_{1}j_{1}}\Lambda^{p_{1}}_{l_{1}i_{1}}=
\varepsilon^{p_{1}}_{k_{1}l_{1}}\Phi^{p_{1}}_{i_{1}j_{1}}
=\Lambda^{p_{1}}_{k_{1}i_{1}}\Lambda^{p_{1}}_{l_{1}j_{1}}-\mathfrak{R}^{m_{1}n_{1}}_{k_{1}l_{1}}
\Lambda^{p_{1}}_{m_{1}j_{1}}\Lambda^{p_{1}}_{n_{1}i_{1}},\\
 \Lambda^{p_{2}}_{k_{2}i_{2}}\Lambda^{p_{2}}_{l_{2}j_{2}}-
 \Lambda^{p_{2}}_{k_{2}j_{2}}\Lambda^{p_{2}}_{l_{2}i_{2}}=
\varepsilon^{p_{2}}_{k_{2}l_{2}}\Phi^{p_{2}}_{i_{2}j_{2}}
=\Lambda^{p_{2}}_{k_{2}i_{2}}\Lambda^{p_{2}}_{l_{2}j_{2}}-\mathfrak{R}^{m_{2}n_{2}}_{k_{2}l_{2}}
\Lambda^{p_{2}}_{m_{2}j_{2}}\Lambda^{p_{2}}_{n_{2}i_{2}},\\
 \Lambda^{p_{3}}_{k_{3}i_{3}}\Lambda^{p_{3}}_{l_{3}j_{3}}-
 \Lambda^{p_{3}}_{k_{3}j_{3}}\Lambda^{p_{3}}_{l_{3}i_{3}}=
\varepsilon^{p_{3}}_{k_{3}l_{3}}\Phi^{p_{3}}_{i_{3}j_{3}}
=\Lambda^{p_{3}}_{k_{3}i_{3}}\Lambda^{p_{3}}_{l_{3}j_{3}}-\mathfrak{R}^{m_{3}n_{3}}_{k_{3}l_{3}}
\Lambda^{p_{3}}_{m_{3}j_{3}}\Lambda^{p_{3}}_{n_{3}i_{3}},
\end{array}
\end{equation}
where $\mathfrak{R}^{m_{1}n_{1}}_{k_{1}l_{1}}=\varepsilon^{m_{1}}_{k_{1}}\varepsilon^{n_{1}}_{l_{1}}$,
$\mathfrak{R}^{m_{2}n_{2}}_{k_{2}l_{2}}=\varepsilon^{m_{2}}_{k_{2}}\varepsilon^{n_{2}}_{l_{2}}$, and
$\mathfrak{R}^{m_{3}n_{3}}_{k_{3}l_{3}}=\varepsilon^{m_{3}}_{k_{3}}\varepsilon^{n_{3}}_{l_{3}}$.

In this paper we have investigate the noncommutative structures of entangled quantum systems. First we have shown that the space of entangled two-qubits can be seen as deformed conifold. Then we wrote the coordinate of this variety in terms of noncommutative space. We have also discussed multipartite entangled systems in terms of noncommutative  geometry.  We  belief that our construction not only important in foundation of quantum theory but it could give rise to new results and applications in the field of quantum information and quantum computing.

\begin{flushleft}
\textbf{Acknowledgments:} This  work was supported  by the Swedish Research Council (VR).
\end{flushleft}


\end{document}